\documentclass[11pt,a4paper]{article}
\usepackage{jcappub}

\pdfoutput=1 

\title{Scalaron dark matter and the thermal history of the universe}

\author{Yuri Shtanov}

\affiliation{Bogolyubov Institute for Theoretical Physics, \\ Metrologichna St.\@ 14-b, Kiev 03143, Ukraine} %
\affiliation{Astronomical Observatory, Taras Shevchenko National University of Kyiv, \\ Observatorna St.\@ 3, Kiev 04053, Ukraine} %

\emailAdd{shtanov@bitp.kyiv.ua}

\abstract{In metric $f(R)$ gravity minimally coupled to the Standard Model, the scalaron field can act as a dark-matter candidate if its mass lies in the range $\text{meV} \lesssim m \lesssim \text{MeV}$. The evolution of the scalaron is influenced by the trace of the stress-energy tensor, whose behaviour, as shown in our previous work, becomes non-adiabatic during the electroweak crossover, potentially triggering scalaron oscillations. While we previously approximated this crossover as a second-order phase transition at the one-loop level, the transition is actually smoother. In this paper, we refine our analysis to account for this smooth crossover and show that scalaron oscillations are still excited in a qualitatively similar manner, driven by the rapid dynamics of the electroweak crossover observed in numerical lattice simulations, provided the scalaron mass is sufficiently small. We also investigate the time-dependent contribution to the stress-energy trace due to the trace anomaly of quantum chromodynamics. Our results indicate that, while the trace anomaly shifts the scalaron's equilibrium value, this shift evolves adiabatically compared to the fast oscillations of the scalaron, meaning that the trace anomaly does not significantly affect the potential cosmological scenarios for scalaron evolution.}


\arxivnumber{2409.05027}

\begin{document} 
\maketitle
\flushbottom

\section{Introduction}

One of the notable dark-matter candidates is the scalar degree of freedom, or scalaron, of $f(R)$ modified gravity \cite{Capozziello:2006uv, Nojiri:2008nt, Cembranos:2008gj, Cembranos:2015svp, Corda:2011aa, Katsuragawa:2016yir, Katsuragawa:2017wge, Yadav:2018llv, Parbin:2020bpp, Shtanov:2021uif, Shtanov:2022xew, KumarSharma:2022qdf} (see also reviews \cite{Sotiriou:2008rp, DeFelice:2010aj}). In the simplest formulation of this model, the $R^2$ term in the series expansion of $f(R)$ plays a crucial role. Here, dark matter consists of a scalaron field oscillating near the minimum of its potential. The scalaron interacts very weakly with matter, which ensures its stability. This model was first proposed in \cite{Cembranos:2008gj} and has been revisited in our recent studies \cite{Shtanov:2021uif, Shtanov:2022xew}. Specifically, we investigated how the scalaron-field oscillations, which constitute dark matter, can be excited and influenced by the electroweak crossover.  

In our previous studies \cite{Shtanov:2021uif, Shtanov:2022xew}, we approximated the electroweak crossover as a second-order phase transition at the one-loop level. However, it is in fact believed to be a smooth crossover without any criticality \cite{Kajantie:1996mn, Bergerhoff:1996jw, Aoki:1996cu, Gurtler:1997hr, Laine:1998jb, Laine:2015kra, DOnofrio:2015gop}. In this paper, we refine our analysis to account for such a smooth crossover and show that scalaron oscillations are excited in a qualitatively similar manner, driven by the rapid dynamics of the electroweak transition observed in numerical lattice simulations.

A significant aspect of our analysis was the use of the local conformal (Weyl) symmetry of the classical action of the Standard Model (SM), except for the Higgs sector, which breaks this symmetry.  However, local conformal symmetry is not only broken classically in the Higgs sector of the Standard Model but also violated due to quantum anomalies. Since the scalaron dynamics is sourced by the trace of the stress-energy tensor of matter, such trace anomalies could potentially impact its evolution. Of particular interest in this respect is the crossover of quantum chromodynamics (QCD), which occurs at a critical temperature of approximately 190~MeV. Previous investigations into the scalaron behaviour in a hot cosmological environment have been conducted using either semi-phenomenological models or toy models describing extensions of the SM (see, e.g., \cite{Katsuragawa:2017wge, Katsuragawa:2018wbe}).\footnote{In a broader context, dynamics of a generic scalar field, mediating a fifth force, in the presence of the cosmological SM thermal bath was recently investigated in \cite{Cyncynates:2024bxw}. Specifically, the authors discuss the role of dimensional transmutation, the QCD scale and the temperature-dependence of the effective potential for the scalar above and below the QCD scale as the SM bath cools down from some unknown reheat temperature. Notably, they observed certain universal characteristics in the scalar's dynamics during the early cosmological epoch.} However, the impact of the trace anomaly of the SM on the scalaron has not been addressed in these studies. This paper aims to fill this gap.

In our earlier studies of the electroweak crossover \cite{Shtanov:2021uif, Shtanov:2022xew}, we found it advantageous to work in the Einstein frame for all fields in the Lagrangian. In this frame, the scalaron classically interacts solely with the Higgs field, simplifying the analysis. However, when addressing the trace anomaly, it is more practical to use the Jordan frame for the matter fields, as this frame preserves the space-time constancy of scales arising in dimensional transmutation, such as the QCD scale \cite{Shtanov:2022wpr}. Therefore, in this paper, we work in the Jordan frame.

We will demonstrate that the trace of the stress-energy tensor influencing the scalaron evolution can be divided into two components: one associated with the Higgs field, which classically breaks local conformal symmetry, and the other related to the rest of the Standard Model, which is classically locally conformally invariant but involves the quantum trace anomaly. The trace anomaly at high temperatures will be addressed qualitatively based on the existing perturbative and lattice simulations. We will show that, given the observational lower bound on the scalaron mass, the trace anomaly has a negligible impact on the scalaron oscillations in the past, as its evolution remains adiabatic relative to the frequency of these oscillations. Consequently, the cosmological scenarios for the scalaron evolution described in \cite{Shtanov:2022xew} remain qualitatively unaffected.

Our paper is organised as follows. In Sec.~\ref{sec:model}, we describe the model of $f (R)$ gravity. In Sec.~\ref{sec:coupling}, we consider the coupling of gravity to matter and derive equations for the scalaron field taking into account the trace anomaly. In Sec.~\ref{sec:higgs}, we review the effect of the electroweak crossover on the scalaron evolution. In Sec.~\ref{sec:anomaly}, we consider the effect of the QCD trace anomaly. Cosmological scenarios for the scalaron dark matter with these effects are considered in Sec.~\ref{sec:scenario}. Our results are summarised In Sec.~\ref{sec:summary}.

\section{The model}
\label{sec:model}

Working in the metric signature $(-, +, +, +)$, we assume that the Lagrangian $L_g$ of an $f (R)$ gravity theory can be expanded in the form of power series in the scalar curvature $R$:
\begin{equation} \label{Sgs}
S_g = \frac{M^2}{3} \int f (R) \sqrt{- g}\, d^4 x \, , \qquad f (R) = - 2 \Lambda + R + \frac{R^2}{6 m^2} + \ldots \, .
\end{equation} 
Here, $M = \sqrt{3 / 16 \pi G} \approx 3 \times 10^{18}\, \text{GeV}$ is a conveniently normalised Planck mass, and $\Lambda \approx \left( 3 \times 10^{-33}\, \text{eV} \right)^2$ is the cosmological constant (corresponding to the vacuum energy density of matter) in the natural units $\hbar = c = 1$.

Proceeding from the Jordan frame to the Einstein frame for the metric, one first writes action \eqref{Sgs} in the form
\begin{equation}\label{Sg1}
S_g = \frac{M^2}{3} \int \bigl[ \Omega R - w (\Omega) \bigr] \sqrt{- g}\, d^4 x \, ,
\end{equation}
where $\Omega$ is a new dimensionless scalar field, and the function $w (\Omega)$ is such that variation with respect to $\Omega$ and its substitution into the action returns the original action. Thus, $f (R)$ is the Legendre transform of $w (\Omega)$, with the direct and inverse transforms being
\begin{align} 
f(R) &= \bigl[ \Omega R - w (\Omega) \bigr]_{\Omega = w'{}^{-1} (R)} \, , \\[3pt]
w (\Omega) &= \bigl[ \Omega R - f (R) \bigr]_{R = f'{}^{-1} (\Omega)} \, , \label{dir2}
\end{align}
which, in principle, enables one to find $w (\Omega)$ given $f (R)$ and vice versa.

Eventually, one can make a conformal transformation in \eqref{Sg1}:
\begin{equation} \label{om}
g_{\mu\nu} = \Omega^{-1} \widetilde g_{\mu\nu} \, , \qquad \Omega = e^{\phi / M} \, ,
\end{equation}
where $\phi$ is a new field (the scalaron) parametrising $\Omega$, and $\widetilde g_{\mu\nu}$ is the metric in the Einstein frame. Action \eqref{Sg1} then becomes that of the Einstein gravity with a canonical scalar field (scalaron) $\phi$ (metric-dependent quantities in the Einstein frame will be denoted by tildes):
\begin{equation}\label{Sg3}
S_g = \int L_g \sqrt{- \widetilde g}\, d^4 x \, , \qquad L_g =  \frac{M^2}{3} \widetilde R - \frac12 \widetilde{\left( \partial \phi \right)}^2 - V (\phi) \, .
\end{equation}
The scalaron potential $V (\phi)$ is calculated by using \eqref{Sg1} and \eqref{dir2}:
\begin{equation} \label{V}
V (\phi) = \frac{M^2}{3} e^{- 2 \phi / M} w \bigl( e^{\phi / M} \bigr) \, .
\end{equation}

Using relation \eqref{dir2}, it is easy to establish that the scalaron potential has extrema, with $V' (\phi) = 0$, at the Jordan-frame values of $R$ that satisfy $R f' (R) = 2 f (R)$. The scalaron mass squared, $m_\phi^2 = V'' (\phi)$, at such an extremum is given by
\begin{equation}
m_\phi^2 = \frac13 \left[ \frac{1}{f''(R)} - \frac{R}{f'(R)} \right] = \frac13 \left[ \frac{1}{f''(R)} - \frac{R^2}{2 f (R)} \right] \, .
\end{equation}
If $m_\phi^2 > 0$, then this is a local minimum. As can be seen from \eqref{dir2}, the scalaron potential varies on a typical scale of the Planck mass $M$.  Hence, in the neighbourhood of the minimum, it is well approximated by a quadratic form on field scales much smaller than $M$.  

For a small cosmological constant, $\Lambda \ll m^2$, the theory has a local minimum at $\phi / M \approx 4 \Lambda / 3 m^2$, corresponding to $R \approx 4 \Lambda$ in the Jordan frame, and the scalaron mass at this minimum is $m_\phi^2 = m^2 + {\cal O} (\Lambda)$. In what follows, we neglect the small cosmological constant in \eqref{Sgs}, which is responsible for dark energy but not for dark matter. In this approximation, the local minimum is situated at $\phi = 0$ (corresponding to $R = 0$ in the Jordan frame), and the scalaron mass at this minimum is $m_\phi = m$. 

As an example, the simplest non-trivial model is described by 
\begin{equation}\label{fstar}
f (R) = R + \frac{R^2}{6 m^2} \, .
\end{equation}
For this model, we have
\begin{equation}\label{w}
w (\Omega) = \frac32 m^2 \left( \Omega - 1 \right)^2 \, , \qquad 
V (\phi) = \frac12  m^2 M^2 \left( 1 - e^{- \phi / M} \right)^2 \, . 
\end{equation}

\section{Coupling to matter and dynamics of the scalaron}
\label{sec:coupling}

The scalaron field satisfies the equation
\begin{equation}\label{eq-phi}
\widetilde \Box\, \phi - V' (\phi) = - \frac{1}{\sqrt{- \widetilde g}} \frac{\delta S_\text{m}}{\delta g^{\mu\nu}} \frac{\delta g^{\mu\nu}}{\delta \phi} = - \frac{e^{- 2 \phi / M}}{\sqrt{- g}} \frac{\delta S_\text{m}}{\delta g^{\mu\nu}} \frac{\delta g^{\mu\nu}}{\delta \phi} = - \frac{e^{- 2 \phi / M}}{2 M} {\cal T}\, ,
\end{equation}
where $S_\text{m}$ is the matter action with the matter fields minimally coupled to the Jordan metric $g_{\mu\nu}$, and ${\cal T}$ is the trace of the stress-energy tensor of matter. This equation can also be written in the form stemming directly from action \eqref{Sg1}:
\begin{equation}\label{eq-om}
\Box\, \Omega - \frac13 \left[ \Omega w' (\Omega) - 2 w (\Omega) \right] = - \frac{1}{2 M^2} {\cal T}\, ,
\end{equation}
with $\Omega (\phi)$ given by \eqref{om}. The right-hand sides of \eqref{eq-phi} or \eqref{eq-om} shift the local equilibrium value of the scalaron field and modify its mass (the so-called `chameleon' effect \cite{Khoury:2003aq, Khoury:2003rn, Brax:2008hh, Burrage:2017qrf, Katsuragawa:2016yir, Katsuragawa:2017wge, Katsuragawa:2018wbe, Katsuragawa:2019uto, Numajiri:2023uif}). An exception is the simplest model \eqref{fstar}, \eqref{w}, for which $\Omega w' (\Omega) - 2 w (\Omega) = 3 m^2 \left( \Omega - 1 \right)$, so that there is no mass `chameleon' effect in Eq.~\eqref{eq-om} \cite{Numajiri:2023uif}. 

As regards the matter part of the action, $S_\text{m}$, we assume that it has the usual form of the Standard Model in the Jordan frame. Note that this action can be split into two parts: $S_\text{\tiny W}$, which is classically Weyl invariant (with appropriate local conformal transformation of the matter fields), and $S_\text{\tiny H}$, which describes the Higgs sector and breaks classical Weyl invariance\footnote{The neutrino sector extending the Standard Model can also fail to be classically Weyl invariant, e.g., if it contains Majorana mass terms. Its contribution is insignificant for the present discussion.}; this part has the field Lagrangian
\begin{equation}\label{Lh}
L_\text{\tiny H} = - \left( D_\mu \Phi \right)^\dagger D^\mu \Phi - \frac{\lambda}{4} \left( 2 \Phi^\dagger \Phi - v^2 \right)^2 \, .
\end{equation}
Here, $D_\mu = \nabla_\mu + A_\mu$ is the gauge covariant derivative involving the SU(2) and U(1) electro\-weak gauge fields $A_\mu$ and acting on the Higgs doublet $\Phi$, $v \approx 246~\text{GeV}$ is the symmetry-breaking parameter, and $\lambda \approx 0.13$ is the self-coupling constant. 

Let us see what this splitting of the action implies for the trace  ${\cal T}$ of the stress-energy tensor on the right-hand side of \eqref{eq-phi} or \eqref{eq-om}. We perform a local conformal transformation of all fields, including the metric field, parametrising this transformation by $\Omega$. Thus, for the metric conformal transformation \eqref{om}, the Higgs field is conformally transformed as $\Phi = \Omega^{1/2}\, \widetilde \Phi$. 
Using the equations of motion (stationarity of the action) for all matter fields, we have (setting $\Omega = 1$ after the variation)
\begin{equation}\label{trans}
\frac{\delta S_\text{m}}{\delta g^{\mu\nu}} \frac{\delta g^{\mu\nu}}{\delta \Omega} = \frac{\delta S_\text{m}}{\delta \Omega} = \frac{\delta S_\text{\tiny H}}{\delta \Omega} + \frac{\delta S_\text{\tiny W}}{\delta \Omega}\, .
\end{equation}
From this relation, we obtain a simple expression for the stress-energy trace including the anomaly:
\begin{equation} \label{T}
{\cal T} \equiv \frac{2}{\sqrt{-g}} \frac{\delta S_\text{m}}{\delta g^{\mu\nu}} g^{\mu\nu} = \frac{2}{\sqrt{-g}} \frac{\delta S_\text{m}}{\delta g^{\mu\nu}} \frac{\delta g^{\mu\nu}}{\delta \Omega}  = {\cal T}_\text{\tiny H} + {\cal T}_\text{an}  \, , 
\end{equation}
where
\begin{equation}\label{Th}
{\cal T}_\text{\tiny H} = \frac{2}{\sqrt{- g}} \left[ \frac{\delta S_\text{\tiny H}}{\delta \Omega} \right]_\text{cl} = \Box\, \bigl( \Phi^\dagger \Phi \bigr) + \lambda v^2 \left(v^2 - 2 \Phi^\dagger \Phi \right)
\end{equation}
is the normal Higgs-field contribution stemming from $S_\text{\tiny H}$, which breaks the classical Weyl invariance, and 
\begin{equation}
{\cal T}_\text{an} = \frac{2}{\sqrt{- g}} \left[ \frac{\delta S_\text{m}}{\delta \Omega} \right]_\text{an}
\end{equation}
is the anomalous trace of the stress-energy tensor. Note that the gauge fields present in \eqref{Lh} did not enter the expression in \eqref{Th}. The details of derivation of \eqref{Th} are presented in Appendix~\ref{sec:app1}.

Note that both components in \eqref{T} depend on the scalaron because they involve the Jordan metric. In what follows, we consider the case where the scalaron is evolving in a close neighbourhood of zero value, as in the scenario where it forms dark matter. In this case, we can retain only the leading terms with respect to the small parameter $\phi / M$. Equation \eqref{eq-phi} or \eqref{eq-om} in this approximation can be written as
\begin{equation}\label{eq-phi-new}
\Box \phi - m^2 \phi = - \frac{1}{2 M} {\cal T}\, ,
\end{equation}
with ${\cal T}$ given by \eqref{T}. The presence of the scalaron degree of freedom in the Jordan metric in this approximation can be neglected.

The structure of the Higgs-field part \eqref{Th} suggests to make a shift of the scalaron field:
\begin{equation}\label{shift}
\phi = \phi' - \frac{\Phi^\dagger \Phi}{2 M} \, .
\end{equation}
Equation \eqref{eq-phi-new} then simplifies to
\begin{equation}\label{eq-phi'}
\Box \phi' - m^2 \phi' = - \frac{1}{2 M} \left({\cal T}'_{\text{\tiny H}} + {\cal T}_\text{an} \right) \, ,
\end{equation}
with
\begin{equation}\label{Th'}
{\cal T}'_{\text{\tiny H}} = {\cal T}_{\text{\tiny H}} + \left( m^2 - \Box \right) \Phi^\dagger \Phi = m^2 \Phi^\dagger \Phi + \lambda v^2 \left(v^2 - 2 \Phi^\dagger \Phi \right) \, .
\end{equation}
This expression does not contain space-time derivatives. 

In \eqref{eq-phi-new}, we neglected the contribution of the stress-energy trace to the scalaron mass (the `chameleon' effect \cite{Khoury:2003aq, Khoury:2003rn, Brax:2008hh, Burrage:2017qrf, Katsuragawa:2016yir, Katsuragawa:2017wge, Katsuragawa:2018wbe, Katsuragawa:2019uto, Numajiri:2023uif}). In this regard, we note that the universal coupling of the scalaron to matter in \eqref{eq-phi-new} produces an additional universal gravitational force of Yukawa type, with the total gravitational potential per unit mass \cite{Stelle:1977ry} 
\begin{equation}
\Phi_\text{grav} = - \frac{2 G}{r} \left( 1 + \frac13 e^{- m r} \right) 
\end{equation}
Non-observation of such additional Yukawa forces between non-relativistic masses at small distances leads to a lower bound on the scalaron mass \cite{Kapner:2006si, Adelberger:2006dh, Murata:2014nra, Perivolaropoulos:2019vkb}
\begin{equation}\label{mlow}
m \geq 2.7~\text{meV} \quad \text{at 95\% C.L.}
\end{equation} 
We will see below that the `chameleon' contribution to the masses of such magnitudes will be negligible at the cosmological epochs of interest.

Coupling of the scalaron to matter allows for the scalaron decays into photons and massive particles provided its mass $m$ is sufficiently large \cite{Cembranos:2008gj, Cembranos:2015svp, Katsuragawa:2016yir, Shtanov:2022xew}. In particular, they would allow for the scalaron decays into electron-positron pairs if $m > 2 m_e$. For the scalaron representing all of dark matter in the universe, an upper bound on its mass was obtained from the observed 511~keV emission line from the Galactic Centre \cite{Cembranos:2008gj, Cembranos:2015svp}:
\begin{equation}\label{mup}
m \lesssim 1.2\, \text{MeV} \, .
\end{equation}

The scalaron decays into photons via the effective anomalous interaction \cite{Cembranos:2008gj, Katsuragawa:2016yir, Shtanov:2022xew}
\begin{equation} \label{pgg}
L_{\phi \gamma \gamma}\, \simeq\, - \alpha_\text{em} \frac{\phi}{M} F_{\mu\nu} F^{\mu\nu} \, ,
\end{equation}
where $\alpha_\text{em} = e^2 / 4 \pi \approx 1/137$ is the fine-structure constant. Given that photons are bosons, one might wonder whether their production could exhibit non-perturbative features, such as parametric resonance, as in the case of the universe preheating \cite{Kofman:1994rk, Kofman:1997yn, Shtanov:1994ce}. However, this does not happen due to the small value of the coupling constant $\alpha_\text{em}$. As a result, the decay of the scalaron remains within the perturbative regime, a point we previously noted in our paper \cite{Shtanov:2021uif} without providing a demonstration. We clarify this issue in Appendix~\ref{sec:app2}.

\section{Effect of the electroweak crossover}
\label{sec:higgs}

In the special case of a spatially homogeneous expanding universe, equations \eqref{eq-phi-new} and \eqref{eq-phi'} read, respectively,
\begin{align} \label{expand}
\ddot \phi + 3 H \dot \phi + m^2 \phi &= \frac{1}{2 M} \left({\cal T}_{\text{\tiny H}} + {\cal T}_\text{an} \right) \, , \\
\ddot \phi' + 3 H \dot \phi' + m^2 \phi' &= \frac{1}{2 M} \left({\cal T}'_{\text{\tiny H}} + {\cal T}_\text{an} \right) \, , \label{expand'}
\end{align}
where $H$ is the Hubble parameter, and an overdot denotes derivative with respect to the cosmological time $t$. 
 
In this section, we review the effect of the Higgs-field part ${\cal T}'_\text{\tiny H}$ on the dynamics of the scalaron field, which we studied in \cite{Shtanov:2021uif, Shtanov:2022xew} in the Einstein frame. In the early hot universe in thermal equilibrium, one should consider thermal average of the right-hand sides in \eqref{expand} and \eqref{expand'}. In this case, the value of the Higgs-field condensate is determined by the temperature-dependent effective potential. Below the electroweak crossover, in one-loop approximation, we have
\begin{equation} \label{PP}
\Phi^\dagger \Phi = \frac{v^2}{2} \left( 1 - \frac{T^2}{T_c^2} \right) \, , \qquad T \lesssim T_c \, , 
\end{equation}
where $T$ denotes the temperature, and $T_c \approx \sqrt{3 \lambda} v \approx 154~\text{GeV}$ is the critical temperature in one-loop approximation. Then, using \eqref{PP}, we estimate the Higgs-field part \eqref{Th'}  as
\begin{equation}\label{Th'est}
{\cal T}'_\text{\tiny H} = \frac{m^2 v^2}{2} \left( 1 - \frac{T^2}{T_c^2} \right) + \lambda v^4 \frac{T^2}{T_c^2} \approx \lambda v^4 \frac{T^2}{T_c^2}  \approx \frac{v^2 T^2}{3}\, , \qquad T \lesssim T_c \, .
\end{equation} 
Above the critical temperature, we have $\Phi^\dagger \Phi \approx 0$ and ${\cal T}_\text{\tiny H} \approx {\cal T}'_\text{\tiny H} \approx \lambda v^4$. The positive contribution from the Higgs-field part of the stress-energy trace to the scalaron mass squared, stemming from \eqref{eq-phi}, is estimated as $\Delta m^2 \leq \lambda v^4 / M^2 \approx 5 \times 10^{- 11}~\text{eV}^2$. This is always negligible given the lower bound \eqref{mlow} on the scalaron mass.

In one-loop approximation, the electroweak transition is formally of second order, and the quantity ${\cal T}'_\text{\tiny H}$ experiences a discontinuity in its time derivative (a velocity kick) at $T = T_c$. Indeed, in this approximation, at $T > T_c$, we have ${\cal T}'_\text{\tiny H} = \lambda v^4 = \text{const}$, so that $\dot {\cal T}'_\text{\tiny H} = 0$, while, at $T < T_c$, according to \eqref{Th'est}, we have $\dot {\cal T}'_\text{\tiny H} \approx - 2 H_c {\cal T}'_\text{\tiny H} = - 2 \lambda H_c v^4$ right below $T_c$. This discontinuity $\Delta \dot {\cal T}'_\text{\tiny H} \approx - 2 \lambda H_c v^4$ in the time derivative is physically instantaneous with respect to the time scale $m^{-1}$ of the scalaron evolution: the scalaron equilibrium position, determined by the right-hand side of \eqref{expand'}, is suddenly set in motion at $T = T_c$. Such a velocity kick could initiate the scalaron oscillations around its equilibrium position, implementing the scalaron dark-matter scenario proposed in our papers \cite{Shtanov:2021uif, Shtanov:2022xew} (see Sec.~\ref{sec:scenario} for more details).\footnote{Triggering the scalaron oscillations by a discontinuity in the stress-energy trace itself at the onset of a first-order electroweak phase transition beyond the Standard Model was discussed in \cite{Katsuragawa:2018wbe}.} 

However, beyond the one-loop approximation, current understanding suggests that the electroweak transition in the Standard Model with the measured Higgs boson mass $m_\text{\tiny H} \approx 125~\text{GeV}$ is not of second order but rather a smooth crossover \cite{Kajantie:1996mn, Bergerhoff:1996jw, Aoki:1996cu, Gurtler:1997hr, Laine:1998jb, Laine:2015kra, DOnofrio:2015gop}. Furthermore, the crossover region cannot be treated perturbatively and necessitates the use of lattice simulations. This highlights the need for further refinement and clarification of the scalaron excitation mechanism proposed in \cite{Shtanov:2021uif, Shtanov:2022xew}. 

Although the electroweak transition is a smooth crossover, it can still proceed rapidly enough across the crossover temperature $T_c$. Evidence for the rapid nature of the electroweak crossover comes from three-loop perturbative calculations combined with lattice simulations of an effective $\text{SU}(2) \times \text{U}(1)$ gauge--Higgs theory \cite{Laine:2015kra, DOnofrio:2015gop}. In particular, the heat capacity in these simulations exhibits a narrow peak, varying over a temperature scale of a few GeV, which is indicative of a rapid crossover (see Fig.~4 of \cite{Laine:2015kra} and Fig.~7 of \cite{DOnofrio:2015gop}). Importantly for our analysis, the complete trace of the stress-energy tensor\,---\,including the anomaly\,---\,exhibits a sharp derivative change at the crossover temperature.

For a light scalaron to be excited to the required degree during the electroweak crossover, it is sufficient that the change $ \Delta \dot {\cal T} $ in the first time derivative of the trace of the stress-energy tensor across the crossover be of the same order as estimated above in the one-loop approximation, i.e., $ \bigl|\Delta \dot {\cal T} \bigr| = \text{a few} \times H_c {\cal T} $, and that this change occur in a time period shorter than the scalaron's oscillation semi-period $ \pi / m $ (thus violating the adiabaticity condition). Such a behaviour is indeed observed in detailed lattice Monte Carlo simulations of an effective $\text{SU}(2) \times \text{U}(1)$ gauge--Higgs theory \cite{DOnofrio:2015gop}, and is, therefore, expected to occur in the full Standard Model. The crossover temperature in these simulations, defined as the peak of the Higgs condensate susceptibility, is found to be $T_c = 159.5 \pm 1.5$~GeV, which is close to the one-loop value. 

\begin{figure}[htp]
\begin{center}
\includegraphics[height=.369\textwidth]{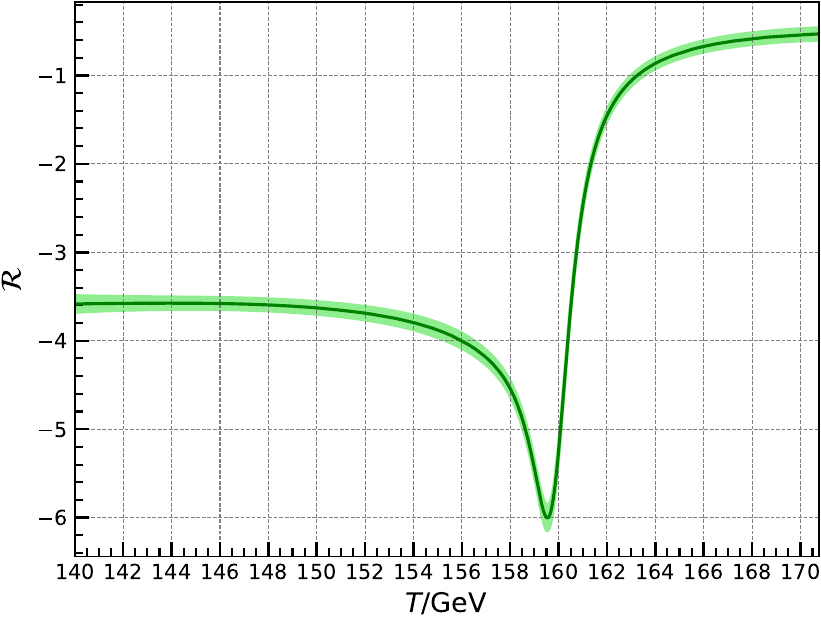}
\includegraphics[height=.369\textwidth]{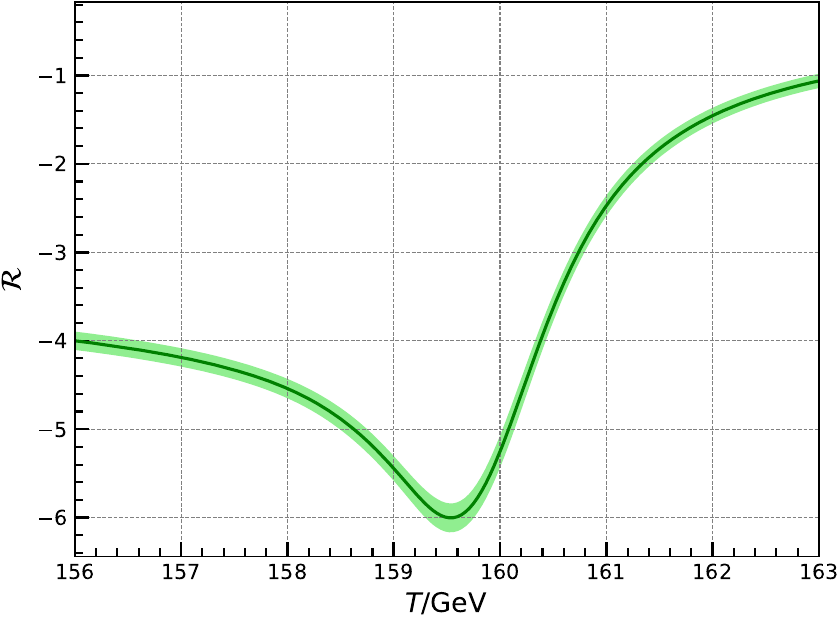}
\caption{Using the Monte Carlo lattice simulation data from \cite{DOnofrio:2015gop}, we plot the dimensionless quantity $ {\cal R} \equiv \dot {\cal T} / H {\cal T} $ as a function of temperature, with the combined statistical error and renormalisation scale variation shown in a colour band. The computed value is affected by a systematic uncertainty in these simulations of order 1\%. The left panel displays the entire available temperature range while the right panel zooms in on the crossover region roughly corresponding to the semi-period $\pi / m$ of oscillations of the scalaron with the minimal mass allowed by \eqref{mlow}.} \label{fig}
\end{center}
\end{figure}

Figure~\ref{fig} shows the evolution of the dimensionless quantity ${\cal R} \equiv \dot {\cal T} / H {\cal T} \approx - d \ln {\cal T} / d \ln T$ that we computed using the publicly available simulation data from \cite{DOnofrio:2015gop}. A relatively sharp drop in the value of ${\cal R}$ is observed in the electroweak crossover, with $\Delta {\cal R} \approx -3$, implying $\Delta \dot {\cal T} \approx -3 H_c {\cal T}$, somewhat larger than in the one-loop approximation. In particular, one can observe a steep drop in ${\cal R}$ from about $- 2.5$ to $- 6$ as the temperature decreases from 161~GeV to $159.5~\text{GeV} = T_c$. In this case, ${\cal T} \approx \lambda v^4$ above the crossover, as in the one-loop approximation, while, below the crossover, the trace ${\cal T}$ decreases faster, as ${\cal T} \propto T^{3.6\, \pm\, 0.1}$, according to Fig.~\ref{fig}.

The Hubble parameter at the crossover temperature is given by
\begin{equation}\label{Hc}
H_c \equiv \left( \frac{\rho_c}{2 M^2} \right)^{1/2} \equiv \left( \frac{\pi^2 g_c T_c^4}{60 M^2} \right)^{1/2} \simeq \left( \frac{3 \pi^2 g_c}{20} \right)^{1/2} \frac{\lambda v^2}{M} \simeq 3 \times 10^{-5}~\text{eV} \, ,
\end{equation}
where $g_c \approx 100$ is the number of relativistic degrees of freedom  in thermal equilibrium at this epoch.  Hence, the semi-period of oscillations of the scalaron with the minimal mass allowed by \eqref{mlow} at this epoch corresponds to a temperature interval $\Delta T = \pi H_c T / m \approx 5.6~\text{GeV}$, which is visually comparable to the width of the crossover in Fig.~\ref{fig}. Given all these parameters, we conclude that excitation of a sufficiently light scalaron (in the meV mass range allowed by \eqref{mlow}) during the electroweak crossover is indeed feasible in the Standard Model. 

A precise numerical evaluation of the excitation degree as a function of the scalaron mass within this model will be conducted in future research. For now, we will limit ourselves to qualitative reasoning. Consider equation \eqref{expand} and denote the deviation of $\phi$ from its equilibrium value by $\chi \equiv \phi - {\cal T} / 2 M m^2$. Equation \eqref{expand} can be written in the form
\begin{equation}\label{chi}
\frac{d}{d t} \left( a^3 \dot \chi \right) + a^3 m^2 \chi = - \frac{1}{2 M m^2} \frac{d}{d t} \left( a^3 \dot {\cal T} \right) \, ,
\end{equation} 
where $a$ is the cosmological scale factor. Now, the field variable $\chi$ together with its time derivative $\dot \chi$ in the scenario under consideration are close to zero prior to the crossover. In the narrow crossover region, the quantity $a^3 \dot {\cal T}$ evolves on a timescale smaller than the oscillation semiperiod $\pi / m$ of a sufficiently light scalaron. Hence, in this region, to a good precision, we can omit the mass term in equation \eqref{chi}. Then this equation relates the time derivative $\dot \chi$ right after the narrow crossover to the variation of $\dot {\cal T}$:
\begin{equation}\label{dot-chi}
\dot \chi \approx - \frac{\Delta \dot {\cal T}}{2 M m^2} \approx - \frac{\Delta {\cal R} H {\cal T}}{2 M m^2} \approx \frac{3 H {\cal T}}{2 M m^2} \, .
\end{equation}
This estimate for $\dot \chi$ immediately after the electroweak crossover is approximately $3/2$ times larger than the prediction from the naive one-loop model \cite{Shtanov:2021uif, Shtanov:2022xew}. Consequently, we can anticipate that the lightest possible scalaron in this scenario will exhibit oscillations with a larger amplitude than in the one-loop approximation.

As temperature drops well below $T_c$, the Higgs-field invariant $\Phi^\dagger \Phi$ approaches its vacuum expectation value $v^2 / 2$ because of contribution from particle masses, and the quantities ${\cal T}'_\text{\tiny H}$ and ${\cal T}_\text{\tiny H}$ then exponentially and adiabatically, relative to the mass scale $m$, approach their vacuum values. Indeed, the leading contributions to the effective potential for the Higgs field during crossover contain exponential factors of the form $e^{- m_i / T}$, where $m_i$ are the (temperature-dependent) particle masses \cite{Laine:2016hma}. The expectation value of the Higgs field at temperatures $T \sim m_i$ will evolve with the rate $t_\text{\tiny H}^{-1} \sim {| \dot T | m_i} / {T^2} \simeq {H m_i} / {T}$. This inverse time scale is at most of the order $H$, decreasing as $t_\text{\tiny H}^{-1} \propto T$ with temperature. Since the product $m t_\text{\tiny H} \sim m / H_c \gg 1$ at the crossover by virtue of \eqref{mlow} and \eqref{Hc}, it remains to be large all the time afterwards. 

At the current cosmological epoch, the component ${\cal T}_\text{\tiny H}$ represents the contribution to the trace of the stress-energy tensor from quarks, leptons and gauge fields whose masses originate from their interactions with the Higgs field. It is remarkable that this contribution can be expressed solely in terms of the Higgs field in \eqref{Th}.

\section{Effect of the QCD trace anomaly}
\label{sec:anomaly}

Let us now examine the effect of the anomalous part ${\cal T}_\text{an}$ in Eq.~\eqref{expand'}. This term arises from the quantum breaking of scale invariance. It includes a vacuum component of order $R^2$ or $\Box R$, which can be ignored in equations \eqref{eq-phi-new} or \eqref{eq-phi'}. The non-vacuum component, however, is significant, as it is thought to be responsible for the masses of hadrons in the Standard Model \cite{Roberts:2022rxm}, making the dominant contribution to the average trace of the stress-energy tensor of matter in the era of colour confinement. The contribution of the total trace to the scalaron mass squared is approximately given by $\Delta m^2 \simeq \rho / M^2$, where $\rho \approx {\cal T}$ is the matter energy density. At the current cosmological epoch, even in the extremely dense objects such as neutron stars, with $\rho \simeq \left( 400~\text{MeV} \right)^4$, this amounts to $\Delta m^2 \simeq 3 \times 10^{- 21}~\text{eV}^2$, which is negligible compared to the lower bound \eqref{mlow}.

At temperatures much higher than the QCD critical temperature $T_\text{\tiny QCD} \simeq 190~\text{MeV}$, one can neglect the baryonic chemical potential, so that the conformal QCD anomaly depends only on temperature and can be estimated as \cite{Kapusta:1979fh, Boyd:1996bx}
\begin{equation}\label{pert}
{\cal T}_\text{an} \simeq \frac{T^4}{\ln^2 T / T_\text{\tiny QCD}} \, . 
\end{equation}
Through the right-hand side of \eqref{eq-phi}, this contributes a positive value $\Delta m^2 = {\cal T}_\text{an}/M^2 \simeq T^4 / \left( M^2 \ln^2 T / T_\text{\tiny QCD} \right)$ to the effective squared mass of the scalaron. However, at high temperatures, where this contribution starts dominating over the scalaron proper squared mass, we have the condition $H^2 \simeq T^4/M^2 \gg \Delta m^2$, so that the scalaron dynamics will be frozen by the Hubble friction in \eqref{expand'} anyway.

Right above the QCD critical temperature, lattice simulations indicate that law \eqref{pert} smoothly passes to an empirical law ${\cal T}_\text{an} \approx 0.3~\text{GeV}^2 T^2$ \cite{Borsanyi:2013bia, HotQCD:2014kol, Bazavov:2015rfa}. Below $T_\text{\tiny QCD}$, this quantity decreases with temperature much faster, but its derivative is still sufficiently small, so that one has $\bigl| \dot {\cal T}_\text{an} / {\cal T}_\text{an} \bigr| \sim H \ll m$. Note that the ratio of the scalaron's oscillation semi-period to the Hubble time at this cosmological epoch is $\pi H / m \lesssim 5 \times 10^{-8}$ even for the lower bound \eqref{mlow} of the scalaron mass. Thus, the cosmological evolution of ${\cal T}_\text{an}$ is always adiabatic with respect to the rapid oscillations of the scalaron. We conclude that the QCD crossover and related trace anomaly do not significantly impact the adiabatic invariant of the scalaron oscillations.

\section{Scenarios for the scalaron dark matter} 
\label{sec:scenario}

As observed in Sec.~\ref{sec:higgs}, the Hubble-friction term becomes weak compared to the scalaron mass term in \eqref{expand'} well before the electroweak crossover. Indeed, the Hubble parameter at the electroweak crossover is given by \eqref{Hc}, and $H_c \ll m$ by virtue of the lower bound \eqref{mlow}. Then, according to Eq.~\eqref{expand'}, the scalar field $\phi'$ at this epoch can oscillate around the central value
\begin{equation}
\bar \phi' = \frac{1}{2 M m^2} \left({\cal T}'_{\text{\tiny H}} + {\cal T}_\text{an} \right) = \frac{\Phi^\dagger \Phi}{2 M} + \frac{\lambda v^2}{2 M m^2} \left(v^2 - 2 \Phi^\dagger \Phi \right) + \frac{1}{2 M m^2} {\cal T}_\text{an} \, .
\end{equation}
Consequently, according to \eqref{shift}, the scalaron $\phi$ at this epoch can oscillate around the value
\begin{equation}
\bar \phi = \bar \phi' - \frac{\Phi^\dagger \Phi}{2 M} = \frac{\lambda v^2}{2 M m^2} \left(v^2 - 2 \Phi^\dagger \Phi \right) + \frac{1}{2 M m^2} {\cal T}_\text{an} \, .
\end{equation}
Prior to the electroweak crossover, we have
\begin{equation}\label{p0}
\bar \phi = \bar \phi' = \frac{1}{2 M m^2} \left( \lambda v^4 + {\cal T}_\text{an} \right) \, .
\end{equation}

In the scenario proposed in \cite{Shtanov:2021uif, Shtanov:2022xew}, prior to the electroweak crossover, the scalaron field follows its equilibrium value $\phi = \phi' \approx \bar \phi'$, as given by \eqref{p0}, without oscillations. In our previous work, we did not include the trace anomaly; thus, $\bar \phi'$ was constant and solved \eqref{expand'} exactly. When including the trace anomaly, $\bar \phi'$ evolves over time. However, the time-derivative terms in \eqref{expand'} can be neglected provided $3 H \ll m$, meaning the scalaron remains in the adiabatic regime. Even considering the lower bound \eqref{mlow} on $m$, this condition translates to $T \ll g^{-1/4}$~TeV, where $g \approx 100$ is the effective number of relativistic degrees of freedom in thermal equilibrium. Thus, the adiabatic regime of the scalaron evolution is established well before the electroweak crossover.

At the electroweak crossover at $T = T_c$, the quantity ${\cal T}'_\text{\tiny H}$ in \eqref{expand'}, estimated in one-loop approximation as \eqref{Th'est}, experiences a rapid velocity kick $\dot {\cal T}'_\text{\tiny H} \approx - 2 H_c {\cal T}'_\text{\tiny H} = - 2 \lambda H_c v^4$, which, according the the described scenario,  triggers oscillations of the scalaron with an initial amplitude $\left( \phi - \bar \phi \right)_a = \left( \phi' - \bar \phi' \right)_a \approx H_c {\cal T}'_\text{\tiny H} / M m^3 = \lambda H_c v^4 / M m^3$, subsequently adiabatically decreasing due to the Hubble friction in \eqref{expand'}. These oscillations form dark matter, and its established amount in the universe in this approximation is obtained if \cite{Shtanov:2021uif, Shtanov:2022xew}
\begin{equation} \label{m}
m \approx 4.4~\text{meV} \, ,
\end{equation}
which is slightly above the lower bound \eqref{mlow}. The quantity \( {\cal T}'_\text{\tiny H} \) evolves adiabatically after the electroweak crossover, while the anomalous trace \( {\cal T}_\text{an} \) always evolves adiabatically, meaning its evolution has no impact on these estimates.

As discussed in Section \ref{sec:higgs}, although our one-loop approximation of the electroweak crossover was imprecise, the crossover observed in numerical lattice simulations of the effective gauge--Higgs theory \cite{Laine:2015kra, DOnofrio:2015gop} occurs rapidly enough that our estimates remain qualitatively accurate for a scalaron in the meV mass range. As follows from our estimate \eqref{dot-chi}, the resulting amplitude of oscillations of a light scalaron in realistic modelling of the crossover may even be somewhat larger than predicted by the one-loop approximation. Therefore, we can expect the value of the required scalaron mass in this scenario to remain similar to or be slightly larger than estimate \eqref{m}, warranting further numerical investigation. 

There is no apparent reason or mechanism that would cause the scalaron to settle at its equilibrium value \eqref{p0} prior to the electroweak crossover. In the original scenario \cite{Cembranos:2008gj, Cembranos:2015svp}, the scalaron field starts out misaligned from its equilibrium value, and the degree of this misalignment and the scalaron mass $m$ then must be finely tuned to match the current dark-matter abundance.  If the initial misalignment is sufficiently large\,---\,or, equivalently, if the scalaron mass is significantly greater than estimate \eqref{m}\,---\,the electroweak crossover and the subsequent evolution of the stress-energy trace ${\cal T}$ have negligible impact on the amplitude of the scalaron oscillations. Consequently, the tuning required in the misalignment scenario remains unchanged \cite{Cembranos:2008gj, Cembranos:2015svp, Shtanov:2022xew}:
\begin{equation}\label{con}
\left( \frac{100}{g_\text{i}} \right)^{1/4} \left( \frac{m}{\text{eV}} \right)^{1/2} \left( \frac{\phi_\text{i} - \bar \phi_\text{i}}{10^{-7}\, M} \right)^2 = \frac{\Omega_\text{\tiny DM} h^2}{0.12} \, .
\end{equation}
Here, the index `i' denotes the time when the field starts oscillating, specifically when $ 3 H \simeq m $. Given the lower bound \eqref{mlow} on $ m $, this corresponds to a temperature $ T_i \gtrsim 2 g_i^{-1/4} $~TeV. The quantity $\Omega_\text{\tiny DM}$ is the cosmological parameter for dark matter, and $h$ is the current Hubble parameter in units of $100$~km/s\,Mpc, with the established value $\Omega_\text{\tiny DM} h^2 = 0.12$ \cite{Planck:2018vyg}. 

In both scenarios, at the onset of the scalaron oscillations, we have the initial condition $\phi_\text{i} \approx \bar \phi_\text{i} \approx \left( {\cal T}_\text{an} \right)_\text{i} / 2 M m^2$, and, using \eqref{p0}, \eqref{pert}, and the condition $3 H_\text{i} \simeq m$,
\begin{equation}\label{cond}
\frac{\phi_\text{i} }{M} \approx \frac{ \left( {\cal T}_\text{an} \right)_\text{i} }{2 M^2 m^2} \simeq \frac{T_\text{i}^4}{2 M^2 m^2 \ln^2 T_\text{i} / T_\text{\tiny QCD}} = \frac{15 H_\text{i}^2}{2 \pi g_\text{i} m^2 \ln^2 T_\text{i} / T_\text{\tiny QCD}} \simeq \frac{1}{\pi g_\text{i} \ln^2 T_\text{i} / T_\text{\tiny QCD}} \ll 1 \, .
\end{equation}

At higher temperatures, we can neglect the mass term in Eq.~\eqref{expand}. Solving this equation with the right-hand side \eqref{pert} and using the relations $\dot T = - H T$ and $H \propto T^2$, we obtain the following asymptotic solution for the scalaron:
\begin{equation}\label{highT}
\frac{\phi}{M} \simeq \alpha + \beta \frac{T}{T_\text{\tiny QCD}} + \frac{\gamma}{\ln T / T_\text{\tiny QCD}} \, , \qquad T \gg T_\text{i} \, , 
\end{equation}
where $\alpha$ and $\beta$ are integration constants, and $\gamma \sim 1/ g_\text{i} \sim 10^{-2}$. The constants $\alpha$ and $\beta$ have to be tuned to ensure the required misalignment condition \eqref{con}.  The second term on the right-hand side of \eqref{highT} describes the so-called decaying mode (it decreases together with $T$) and may be suppressed in the early universe in which inflation took place. 

Constraints on the inflationary energy scale, derived from the requirement of adiabaticity in the primordial dark-matter perturbations, remain unchanged even after including the trace anomaly, as they depend solely on the deviation of the scalaron from its equilibrium value. For the inflationary Hubble parameter, these constrains read \cite{Shtanov:2022xew}
\begin{equation}
H_\text{inf} \, \lesssim \, 3 \times 10^6\, \left( \frac{g_\text{i}}{100} \right)^{1/8} \left( \frac{\text{eV}}{m} \right)^{1/4}~\text{GeV} \, ,
\end{equation}
implying an upper bound on the reheating temperature, 
\begin{equation}
T_r \, \lesssim \, 10^{12} \left( \frac{\text{eV}}{m} \right)^{1/8}~\text{GeV} \, .
\end{equation}

\section{Summary}
\label{sec:summary}

In this paper, we revisited the cosmological scenario of production of dark matter in the form of the scalaron of $f(R)$ gravity minimally coupled to the Standard Model \cite{Cembranos:2008gj, Cembranos:2015svp, Shtanov:2021uif, Shtanov:2022xew}. Specifically, we refined the mechanism of the scalaron excitation during the electroweak crossover proposed in \cite{Shtanov:2021uif, Shtanov:2022xew} and examined the impact of the trace anomaly at high temperatures in the early universe. We have shown that excitation of scalaron oscillations can be driven by the rapid nature of the electroweak crossover, confirmed by numerical lattice simulations \cite{Laine:2015kra, DOnofrio:2015gop}, provided the scalaron mass is in the meV range. We have also demonstrated that, while the trace anomaly shifts the central value around which the scalaron oscillates, this shift evolves adiabatically with respect to the rapid scalaron oscillations. Consequently, the cosmological scenarios described in \cite{Shtanov:2022xew} remain qualitatively unaffected by these corrections. Unlike our previous work \cite{Shtanov:2021uif, Shtanov:2022xew}, which was conducted in the Einstein frame, this analysis was performed in the Jordan frame, once again demonstrating the consistency of using different frames.

To form dark matter, the scalaron must evolve in a region very close to its zero value. At the time it starts oscillating, this requirement is expressed by constraint \eqref{con} and condition \eqref{cond}. During periods of higher temperatures, where the scalaron is frozen and does not oscillate, its evolution is governed by the trace anomaly, as described by \eqref{highT}. In this scenario, the primordial value of the scalaron appears to be unpredictable, suggesting that the anthropic principle might need to be invoked in models where the scalaron constitutes dark matter \cite{Shtanov:2022xew}.

\section*{Acknowledgements}

I am indebted to my colleague Anton Rudakovskyi for his assistance in preparing the figures. I am also grateful to the anonymous referee for their stimulating comments and constructive feedback. This research was funded by the National Academy of Sciences of Ukraine under project 0121U109612, by the National Research Foundation of Ukraine under project 2023.03/0149, and by the Simons Foundation.

\appendix

\section{Higgs-field contribution to the stress-energy trace}
\label{sec:app1}

We give some details of derivation of the Higgs-field contribution \eqref{Th} to the stress-energy trace stemming from \eqref{trans}. Consider the kinetic term in \eqref{Lh}. Performing a local conformal transformation of fields with parameter $\Omega$ (specifically, $g_{\mu\nu} \to \Omega^{-1} g_{\mu\nu}$ and $\Phi \to \Omega^{1/2}\, \Phi$), we obtain the corresponding variation of this part of the action at $\Omega = 1$ (the arrows indicate the direction of derivatives):
\begin{align}\label{kin}
- \delta \int \left( D_\mu \Phi \right)^\dagger D_\nu \Phi\, g^{\mu\nu} \sqrt{- g}\, d^4 x = - \frac12 \int \bigl( \delta \Omega\, \Phi^\dagger \bigr) \left( \overleftarrow \nabla_\mu + A_\mu^\dagger \right) \left( \overrightarrow \nabla_\nu + A_\nu \right) \Phi \, g^{\mu\nu} \sqrt{- g}\, d^4 x \nonumber \\[3pt] 
- \frac12 \int \Phi^\dagger \left( \overleftarrow \nabla_\mu + A_\mu^\dagger \right) \left( \overrightarrow \nabla_\nu + A_\nu \right) \bigl( \delta \Omega\, \Phi \bigr) \, g^{\mu\nu} \sqrt{- g}\, d^4 x \nonumber \\[3pt] 
+ \int \Phi^\dagger \left( \overleftarrow \nabla_\mu + A_\mu^\dagger \right) \left( \overrightarrow \nabla_\nu + A_\nu \right) \Phi \, \delta \Omega\, g^{\mu\nu} \sqrt{- g}\, d^4 x \nonumber \\[3pt] 
= \frac12 \int \delta \Omega \left\{ \overrightarrow \nabla_\mu \left[ \Phi^\dagger \left( \overrightarrow \nabla_\nu + A_\nu \right) \Phi \right] + \overrightarrow \nabla_\nu \left[ \Phi^\dagger \left( \overleftarrow \nabla_\mu + A_\mu^\dagger \right) \Phi \right] \right\} \, g^{\mu\nu} \sqrt{- g}\, d^4 x \nonumber \\[3pt] 
= \frac12 \int \delta \Omega\, \overrightarrow \nabla_\mu \left[ \Phi^\dagger \left( \overrightarrow \nabla_\nu + \overleftarrow \nabla_\nu \right) \Phi \right]  \, g^{\mu\nu} \sqrt{- g}\, d^4 x = \frac12 \int \delta \Omega\, \Box\, \bigl( \Phi^\dagger \Phi \bigr)  \, \sqrt{- g}\, d^4 x \, .
\end{align}
Here, we have used integration by parts and the relation $A_\mu^\dagger + A_\mu = 0$, valid in our notation. Expression \eqref{kin} gives the kinetic term in \eqref{Th}. The contribution to \eqref{Th} from the potential term in \eqref{Lh} is easily calculated in a similar way. 

\section{Absence of parametric resonance in the production of photons}
\label{sec:app2}

Consider the following interaction of a time-dependent field $I (t)$ with electromagnetism:
\begin{equation} \label{Lem}
L_I = - \frac14 I^2 F_{\mu\nu} F^{\mu\nu} \, .
\end{equation}
This type of interaction is widely used in the theory of inflationary magnetogenesis. Adopting the longitudinal gauge $A_0 = 0$, $\partial^i A_i = 0$ for the vector potential, from (\ref{Lem}) one obtains the equation satisfied by the transverse field variable $A_i$ in the spatial Fourier representation (see, e.g., \cite{Demozzi:2009fu, Shtanov:2020gjp}):
\begin{equation} \label{eqAi}	
{\cal A}_i''  + \left( k^2 - \frac{I''}{I} \right) {\cal A}_i = 0 \, , 
\end{equation}
where ${\cal A}_i = I A_i$, $k$ is the comoving wavenumber, and the prime denotes the derivative with respect to the conformal cosmological time $\eta$, related to the physical time $t$ by $d \eta = d t / a$, where $a$ is the cosmological scale factor. 
If the function $I$ is evolving quasi-periodically, then we have a classical problem of parametric resonance. 

In our case of the scalaron interaction \eqref{pgg}, we have
\begin{equation} \label{I}
I = 1 + 2 \xi \frac{\phi}{M} \, , \qquad - \frac{I''}{I} \simeq 2 \xi m^2 a^2 \frac{\phi}{M} \, , 
\end{equation}
where $\xi \sim \alpha_\text{em}$ denotes a numerical constant, and we have used the fact that $\alpha_\text{em} \ll 1$, $|\phi|/M \ll 1$, and $H \ll m$. The general conditions on the strength of interaction under which the parametric resonance develops in the expanding universe are well known \cite{Kofman:1994rk, Kofman:1997yn, Shtanov:1994ce}. For the coupling \eqref{I}, the parametric resonance will be narrow since $\xi |\phi| / M \ll 1$. In this case, the Floquet exponent $\mu$ with respect to the physical time $t$ is given by \cite{Shtanov:1994ce}
\begin{equation} \label{mu}
\mu = \frac{1}{m} \sqrt{\frac{\xi^2 m^4 \phi_a^2}{M^2} - \left( \frac{k^2}{a^2} - \frac{m^2}{4} \right)^2 } \, ,
\end{equation}
where $\phi_a$ is the (time-dependent) amplitude of the scalaron oscillations. This Floquet exponent describes both the resonance bandwidth and amplification of the electromagnetic mode in time, ${\cal A}_i \propto \exp \int \mu d t$. However, the exponential resonant amplification takes place only if the Floquet exponent itself varies adiabatically with time: $| \dot \mu / \mu| \ll \mu$, which, in view of \eqref{mu}, requires the condition $H \ll \mu$. In the opposite case $\mu \ll H$, the universe expansion quickly redshifts the modes out of the narrow resonance band preventing the resonance from developing, and the field excitation proceeds according to the ordinary perturbation theory, as discussed in detail in \cite{Rudenok:2014daa}. The condition $\mu \ll H$, in view of \eqref{mu}, reads
\begin{equation}
\xi \frac{m \phi_a}{M} \ll H \, .
\end{equation}
Taking into account that $H^2 = \rho / 2 M^2$, where $\rho$ is the total energy density (including dark energy), and $m^2 \phi_a^2 = 2 \rho_\phi$, where $\rho_\phi$ is the energy density of scalaron oscillations, we finally arrive at the condition \cite{Shtanov:2021uif}
\begin{equation}
\alpha_\text{em} \sqrt{\frac{\rho_\phi}{\rho}} \ll 1 \, ,
\end{equation}
which is satisfied due to the smallness of $\alpha_\text{em}$ with respect to unity and because $\rho_\phi / \rho < 1$.

Note that, in the early hot universe, electromagnetic field is not free but is in tight thermal equilibrium with the rest of matter, and its resonant amplification cannot occur for this reason alone. Equation \eqref{eqAi} in this case is modified to an effective equation (see \cite{Shtanov:2020gjp})
\begin{equation} 	\label{eqAs}
{\cal A}_i''  + \frac{a \sigma}{I^2} {\cal A}_i' + \left( k^2 - \frac{I''}{I} - \frac{a \sigma I'}{I^3} \right) {\cal A}_i = 0 \, , 
\end{equation}
where $\sigma$ is the electric conductivity of the primordial plasma. This conductivity is huge, $\sigma \sim T / \alpha_\text{em} \ln \left( 1 / \alpha_\text{em} \right) \gtrsim 30\, T$ \cite{Baym:1997gq}, and the dissipation term in \eqref{eqAs} is sufficient to kill the resonance.

In the regions of virialised dark-matter halos, the Floquet exponent is estimated as
\begin{equation}\label{mut}
\mu \simeq \xi \frac{m \phi_a}{M} \sim \alpha_\text{em} \frac{\sqrt{ \rho_\phi}}{M} \simeq \alpha_\text{em} \sqrt{ \frac{\rho_\phi}{\bar \rho}} H \, ,
\end{equation} 
where $\rho_\phi$ is the local scalaron energy density, and $\bar \rho$ is the mean cosmological energy density (including dark energy). Parametric resonance in the scalaron decay into photons could take place only if the scalaron oscillations were coherent on space-time scales larger than $\mu^{-1}$, which, in view of \eqref{mut}, is comparable to the Hubble radius. However, for the scalaron forming dark matter, its coherence length is of the order of the de~Broglie wavelength $\lambda_\text{dB}$, which is much smaller than the halo size that, in turn, is much smaller than the Hubble radius. For instance, at the present cosmological epoch, it is estimated as \cite{Shtanov:2022wpr}
\begin{equation}
\lambda_\text{dB}\, \lesssim\, \frac{10^4}{m} \simeq 10^5 \left( \frac{\text{meV}}{m} \right)\,\text{cm}\, \lll\, H_0^{-1} \simeq 10^{28}\, \text{cm} \, . 
\end{equation}
Thus, parametric resonant amplification of electromagnetic field in dark-matter halos is also prohibited, and the scalaron creates photons according to the ordinary perturbation theory \cite{Shtanov:2021uif}.

\bibliographystyle{JHEP}
\bibliography{scalaron}

\end{document}